\documentclass[letterpaper]{mc2025}

\usepackage{algorithm}
\usepackage{algorithmic}

\usepackage{nomencl} 
\makenomenclature

\title{A Batch Power Iteration Approach for the Iterative Quasi-Monte Carlo Method Using a Randomized-Halton Sequence}

\addAuthor[sam.pasmann@thea.energy]{Samuel Pasmann}{1}
\addAuthor{Ilham Variansyah}{2}
\addAuthor{C.T. Kelley}{3}
\addAuthor{Ryan G. McClarren}{4} 

\addAffiliation{1}{Thea Energy, Kearny, NJ}
\addAffiliation{2}{Oregon State University, Corvallis, OR}
\addAffiliation{3}{North Carolina State University, Raleigh, NC}
\addAffiliation{4}{University of Notre Dame, Notre Dame, IN}

\Abstract{The Iterative Quasi-Monte Carlo (iQMC) method is a recently developed hybrid method for neutron transport simulations. iQMC replaces standard quadrature techniques used in deterministic linear solvers with Quasi-Monte Carlo simulation for accurate and efficient solutions to the neutron transport equation. Previous iQMC studies utilized a fixed-seed approach wherein particles were reset to the same initial position and direction of travel at the start of every transport sweep. While the QMC samples offered greatly improved uniformity compared to pseudo-random samples, the fixed-seed approach meant that some regions of the problem were under-sampled and resulted in errors similar to ray effects observed in discrete ordinates methods.

This work explores using randomized-Quasi Monte Carlo techniques (RQMC) to generate unique sets of QMC samples for each transport sweep and gain a much-improved sampling of the phase space. The use of RQMC introduces some stochastic noise to iQMC's iterative process, which was previously absent. To compensate, we adopt a ``batch'' approach similar to typical Monte Carlo k-eigenvalue problems, where the iQMC source is converged over $N_\text{inactive}$ batches, then results from  $N_\text{active}$ batches are recorded and used to calculate the average and standard deviation of the solution.

The RQMC batch method was implemented in the Monte Carlo Dynamic Code (MC/DC) and is shown to be a large improvement over the fixed-seed method. The batch method was able to provide iteratively stable and more accurate solutions with nearly two orders of magnitude reduction in the number of particle histories per batch. Notably, despite introducing some stochastic noise to the solution, the RQMC batch approach converges both the k-effective and mean scalar flux error at the theoretical QMC convergence rate of $O(N^{-1})$.
}


\shortTitle{A RQMC Batched Power Iteration for iQMC}

\authorHead{S. Pasmann, et al.}

\begin{document}

\section{INTRODUCTION}\label{sec:introudction}

\subsection{The Iterative Quasi-Monte Carlo Method for Neutron Transport}
The iterative-Quasi-Monte Carlo (iQMC) method is a recently developed hybrid method for multigroup neutron transport simulations \cite{Pasmann2023Quasi}. iQMC is the combination of Monte Carlo (MC) simulation, deterministic iterative methods, and quasi-Monte Carlo (QMC) techniques. iQMC approximates solutions to the neutron transport equation using successive QMC transport sweeps. QMC transport sweeps are reminiscent of analog MC simulation, in that particles are tracked continuously in space/angle and can therefore be tracked on the same arbitrarily complex geometries. 

Unlike analog MC, iQMC does not explicitly model the scattering and fission processes. Instead, iQMC treats the scattering and fission processes as internal fixed sources. The scattering and fission sources are discretized across the domain using either a piecewise-constant or piecewise-linear source approximation stored in a global mesh~\cite{pasmann2024mitigating}. When iQMC particles are initialized, their statistical weight is proportional to the source ``strength'' of the scattering and fission terms at the emission site. The particles are then traced out of the volume, attenuating the statistical weight with continuous weight absorption and tallying the cell-averaged scalar flux and effective source (scattering and fission) with a path-length tally estimator. By removing the need explicitly to model the scattering and fission processes, we are left with a purely absorbing system, and consequently, the iQMC transport sweep is reduced to a ray-trace operation. After $N$ particle sweeps, we arrive at an updated estimate of the flux and our source terms and can proceed in an iterative fashion.

Given the global fixed source, particles need to be sampled uniformly across the domain. This requirement for global uniform sampling and ray-trace procedure provides a well-suited application for QMC sampling. QMC is the replacement of pseudo-random number generators in Monte Carlo with low-discrepancy sequences (LDSs). The LDSs use quasi-random or deterministic algorithms to generate sequences with maximum distances between samples. This results in a more efficient sampling of the phase-space and, for $N$ samples, a theoretical $O(N^{-1})$ convergence, compared with the $O(N^{-1/2})$ convergence rate from analog Monte-Carlo [4]. In iQMC's purely absorbing ray-trace, this means QMC samples are only used to sample a particle's initial position and direction of travel.

In addition to an improved convergence rate compared to analog MC, iQMC holds several algorithmic characteristics that may be advantageous in high-performance computing environments, including:
\begin{itemize}
    \item A vectorized multigroup scheme where each particle can represent all energy groups.
    \item A non-divergent algorithmic scheme, \textit{i.e.} the ray-trace avoids embedded conditional statements - a promising feature for GPU computations.
    \item Sampling particles uniformly across the domain results in a more geometrically balanced workload, an advantage for domain-decomposed simulations.
    \item Good approximations of the flux even in regions of very low flux.
    \item  Continuous angular treatment, an ability to handle complex 3D geometries, and a highly parallel nature similar to MC simulation.
\end{itemize}

While iQMC has many computational advantages, it also has its drawbacks. Mainly, the global mesh, required to store the piecewise source approximation, introduces some amount of spatial discretization error to the solution. It has been shown that refining the mesh and/or using a piecewise linear rather than a piecewise constant source approximation can reduce this error~\cite{pasmann2024mitigating}. It has also been observed that iQMC requires relatively large particle counts to adequately resolve iQMC's linear source term~\cite{pasmann2024mitigating}. This problem likely stems from iQMC's originally formulated fixed-seed approach, where particles are reset to their original position and direction of travel at the start of each transport sweep. This work investigates the use of randomized QMC sampling in a new iterative formulation of the iQMC method and evaluates the impact on numerical performance relative to the fixed-seed approach.

\subsection{QMC Fixed-Seed versus Randomized-QMC Batching}

Although LDS, like the Halton sequence~\cite{halton1960efficiency}, Sobol sequence~\cite{sobol1967distribution}, and Latin hyper-cube~\cite{mckay2000comparison} provide very uniformly distributed samples, the sequences themselves are deterministic. Each point generated in a sequence is dependent on the previous, and it's inadvisable in QMC to split the sequence and use separate chunks as individual sample sets. Therefore, in the context of MC integration, the only way to further explore the phase space is by taking more samples from the sequence (as opposed to pseudo-random samples where an arbitrary number of sample sets can be created). In iQMC, this translates to more particle histories per transport sweep. Each transport sweep will run through the same $N$ samples across $D$ dimensions (typically five: three in space and two in angle). Therefore, particles are emitted at the same position and travel in the same direction as they did in the previous sweep. This is equivalent to setting a fixed random number seed before each batch in analog MC simulation. 

This technique is referred to as the ``fixed-seed'' approach and comes with some advantages. Namely, by resetting particles to the beginning of the sequence at the start of each sweep, the only change in the flux per iteration is a result of the numerical convergence towards the final converged solution. Importantly, this allows the QMC transport sweeps to be used as matrix-vector product functions that uphold a linearity assumption -- a necessary condition for using advanced linear Krylov solvers. Linear Krylov solvers like generalized minimal residual (GMRES) for fixed source problems and the generalized Davidson method for k-eigenvalue problems have been shown to converge with far fewer iterations in iQMC than the typical source iteration or power iteration~\cite{Pasmann2023iQMC, Pasmann2021}. However, by fixing the seed, some regions of the problem are never sampled, and we are introducing some amount of bias to the solution, similar to the beam-like artifacts, or ``ray effects'', in discrete ordinates methods.

The alternative to a fixed seed approach would be a batch approach similar to what is used in typical MC k-eigenvalue simulations~\cite{McClarren2018} and The Random Ray Method~\cite{Tramm2017}. Rather than resetting the particles to the same position, a batched approach would emit particles at new locations, traveling in new directions at the start of every transport sweep. This means the LDS needs to be scrambled or randomized using randomized-QMC (RQMC) techniques at the start of every transport sweep. RQMC techniques have seen increasing interest and research over the last few decades~\cite{lecuyer2018Randomized} and are designed to construct each point individually $U\left[0,1\right]^D$ while collectively the $N$ points retain their low discrepancy. By randomizing the sequence, we can generate a unique set of (R)QMC samples for each transport sweep. This would allow for a much more extensive sampling of the phase space over the course of $R$ transport sweeps, and the number of particle histories required for an accurate solution approximation may be significantly reduced.

However, by randomizing the QMC samples, we are also introducing some amount of stochastic noise to the solution. We can mitigate this noise in the same way that many MC k-eigenvalue algorithms do, by splitting the simulation into ``inactive'' and ``active'' transport sweeps or ``batches''. In iQMC, inactive batches iterate on the scalar flux and source strength until the relative change between iterations becomes stationary. Then active batches begin, where tally histories are recorded and later used to calculate the mean and standard deviation of each tally estimate. In fixed-seed iQMC, samples are deterministic, which makes it difficult to estimate the accuracy of the converged solution. While the batched RQMC approach has several advantages over the fixed-seed, it would also destroy the linearity assumption of our matrix-vector product functions and, therefore, negate the use of linear Krylov solvers (GMRES and Davidson's method).

\section{iQMC Batch Power Iteration Algorithm}\label{sec:methodology}
The first important consideration in designing a batched iQMC mode is in the selection of the RQMC method. As with QMC methods, there are a host of RQMC methods, which have been developed over the last several decades~\cite{lecuyer2002recent}. The method chosen for this work was Owen's randomization of the Halton Sequence~\cite{Owen2017}. The Halton sequence is itself a multidimensional extension of the van der Corput sequence~\cite{halton1960efficiency}, which is extensible both in $N$ and $D$. Importantly, Owen's randomization uses independent random permutations to randomize the samples and, therefore, can be called an arbitrary number of times to generate $R$ unique sample sets. While the Halton sequence is typically not as accurate as the Sobol sequence, the Halton sequence is much less sensitive to the number of samples $N$ generated in the set whereas Sobol sequences are built to produce samples of $N$ in powers of 2~\cite{sobol1967distribution}. Additionally, Owen's randomization is particularly useful in applications like iQMC because it can generate $N^\prime$ new samples by only generating the $\left[N,N^\prime\right]$ rows -- an important feature for generating samples in parallel computations. Similar to traditional MC simulation, the samples produced by Owen's randomization can be reproduced by passing a seed to the local pseudo-random number generator used for randomizing the samples. Finally, Owen's randomization was chosen due to its relative ease of implementation, and it does not require any additional input from the user.

With a chosen RQMC method, it is relatively straightforward to adapt the fixed-seed iQMC flattened power iteration for k-eigenvalue problems~\cite{Pasmann2023iQMC} to incorporate the new sampling method and batch iteration scheme. Unlike the fixed-seed method, the change in global variables in the batch method will be due to convergence towards the final solution \textit{and} some stochastic noise introduced from RQMC samples. Consequently, there must be enough inactive batches to allow for the source strength to stabilize and avoid introducing bias to the averaged solution. Similarly, there need to be enough active batches to converge the solution and mitigate the stochastic noise. Ideally, both $N_\text{inactive}$ and $N_\text{active}$ values would be determined by some quantitative metric, like the relative difference in the source strength per batch. For now, however, these values are input from the user, and selecting the right number of inactive/active batches can be approximated from studying the convergence of the solution in each problem.

For each new inactive and active batch, we generate a new set of RQMC samples to initialize particles. After generating the samples, the power iteration proceeds as normal with the QMC transport sweep and update of global variables, like k-effective. A general outline of this algorithm is shown in Algorithm~\ref{alg:batch_pi}. As mentioned above, the inactive batch phase should iterate long enough until the noise between iterations stabilizes, then the active batch phase begins, and we begin to accumulate an average of the global variables after each sweep.

\begin{minipage}{0.7\textwidth}
\begin{algorithm}[H]
    \caption{iQMC Batch Power Iteration}
    \label{alg:batch_pi}
    \begin{algorithmic}[1]
        \FOR {$(N_\text{inactive}+N_\text{active})$ batches}
            \STATE Generate RQMC samples with Owen's Randomization
            \FOR {$N_\text{particles}$}
                \STATE Initialize particle position and angle from the RQMC samples
                \STATE Initialize particle weight from the effective source tally
                \WHILE {Particle is alive}
                    \STATE Calculate distance to next boundary
                    \STATE Tally: scalar flux, effective source, and fission source
                    \STATE Attenuate particle weight
                    \STATE Advance particle
                \ENDWHILE
           \ENDFOR
            \STATE Update $k_\text{eff}$ with new estimate of the fission source
            \IF {batch number $> N_\text{inactve}$}
                \STATE Record tally scores
            \ENDIF
        \ENDFOR
        \STATE Return mean and standard deviation of tally scores from $N_\text{active}$ batches
    \end{algorithmic}
\end{algorithm}
\end{minipage}

\subsection{Implementation in the Monte Carlo Dynamic Code}
iQMC has been implemented and verified in the Monte Carlo Dynamic Code (MC/DC), a performant, scalable, and machine-portable Python-based Monte Carlo neutron transport software currently in development in the Center for Exascale Monte-Carlo for Neutron Transport (CEMeNT)~\cite{morgan2024mcdc}. MC/DC was designed for rapid prototyping of neutron transport algorithms and takes advantage of \textit{Numba}~\cite{lam2015numba}, a just-in-time compiler for scientific computing in Python for increased performance. MC/DC also utilizes mpi4py~\cite{rogowski2023mpi4py} for parallel computation across multiple cores.

\begin{figure}[h]
  \centering
  \includegraphics[width=\textwidth]{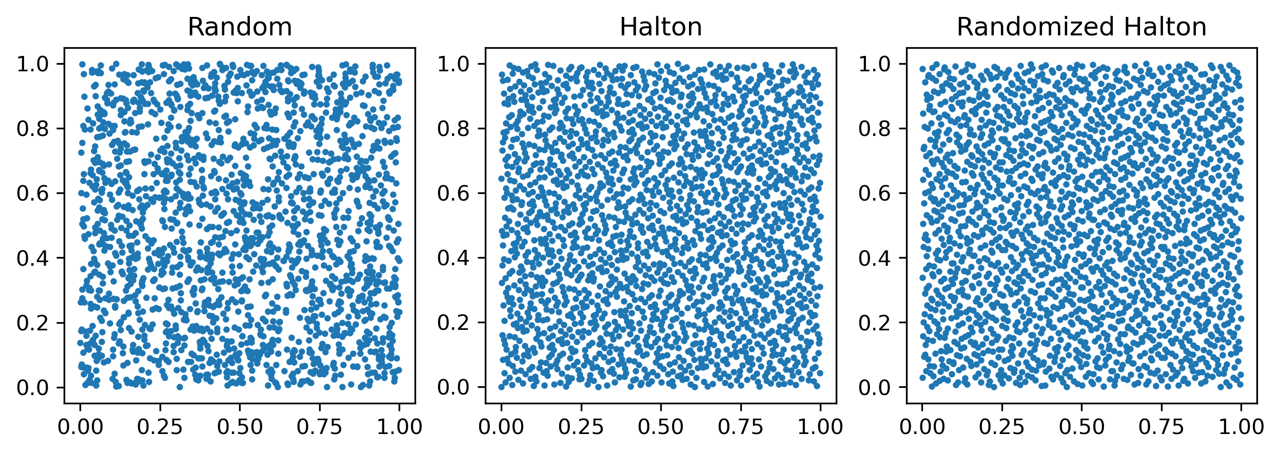}
  \caption{Points generated in the unit square from a pseudo-random number generator, Halton Sequence, and randomized Halton sequence.}
  \label{fig:samples}
\end{figure}

In iQMC's first implementation in MC/DC, each MPI rank pre-computed a portion of the low-discrepancy sequence using Scipy's QMC package~\cite{2020SciPy-NMeth} before entering the JIT compiled section of the code. Each rank stored this matrix of size $N \times 5$ (five dimensions per particle: three in space and two in angle) and would reference these stored samples each transport sweep to reset the particles. Scipy offers unscrambled and scrambled versions of the Halton sequence but randomizing the sequence after each transport sweep in the batched approach would require exiting the compiled code with Numba's \texttt{objmode}, each time the function is called -- a costly routine. To avoid this, both the scrambled and unscrambled implementations of the Halton sequence were implemented directly in MC/DC. Figure~\ref{fig:samples} shows samples generated from a pseudo-random number generator, the unaltered Halton sequence, and the Randomized Halton sequence using MCDC's custom implementation of the Halton and Randomized Halton algorithms.

\section{TAKEDA-1 RESULTS}\label{sec:results}
This section introduces the first preliminary numerical experiments with iQMC's batched RQMC method using the Takeda-1 benchmark problem. Figure~\ref{fig:takeda} shows the Takeda-1 benchmark a simplified, 3D, 2-group, k-eigenvalue reactor problem~\cite{Takeda1991}. The Takeda-1 problem features three regions: core, control rod, and moderator. There are two variations of the problem: one is where the control rod region is replaced by void, and the other is where the control rod is inserted so that the region is highly absorbing. The results presented were run with the control rod inserted. iQMC and analog Monte Carlo reference results were generated on a $25\times25\times25$ regular grid. The reference eigenvalue and scalar flux solution were generated with analog multigroup Monte Carlo in MC/DC with 50 million particle histories per batch with 10 inactive and 20 active batches.
\textbf{\begin{figure}[ht]
  \centering
  \includegraphics[width=\textwidth]{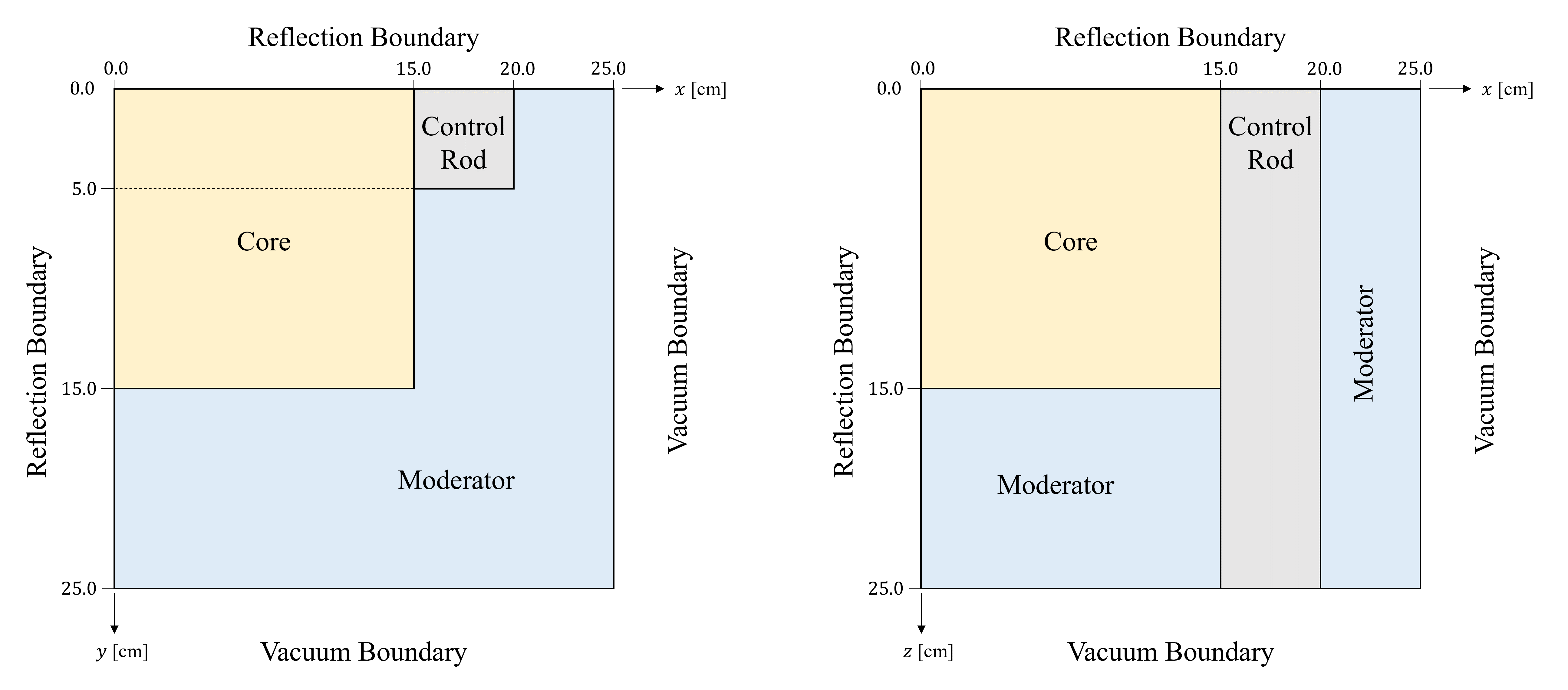}
  \caption{The Takeda-1 k-Eigenvalue benchmark problem~\cite{Takeda1991}.}
  \label{fig:takeda}
\end{figure}}

The first experiment, shown in Figure~\ref{fig:batched_and_fixed_seed}, compares k-Effective after each transport sweep/batch for one simulation using the fixed-seed and batch methods. Both methods utilized iQMC's more accurate piecewise-linear source approximation, but a relatively high number of particle histories ($N=1.5E6$ per transport sweep) were required to keep the fixed-seed solution from diverging. The batch method was able to produce a stable solution with far fewer particle histories per transport sweep (about two orders of magnitude fewer), a point which is emphasized in Figure~\ref{fig:halton_and_random_batches} and~\ref{fig:error}. It's immediately obvious from Figure~\ref{fig:batched_and_fixed_seed} that the batch RQMC k-effective approximation is significantly more accurate than the fixed-seed approximation. The accuracy of the fixed-seed approach is limited by the rays traced by the $N=1.5E6$ source particles that are identically emitted throughout the iteration. The accuracy of the fixed-seed approach improves as $N$ increases. We also note that the batch k-effective does not appear to stabilize until after approximately 200 transport sweeps, indicating this is the approximate number of inactive batches that should be run before switching to active batches and collecting batch statistics.
\textbf{\begin{figure}[ht]
  \centering
  \includegraphics[width=0.8\textwidth]{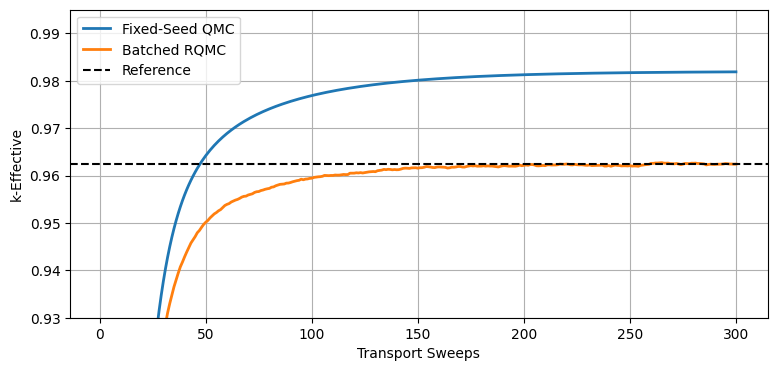}
  \caption{Updated k-Effective approximation after each transport sweep from the Takeda-1 problem with the iQMC fixed seed and batched approach. Each simulation was run using 1.5E6 particles per transport sweep.}
  \label{fig:batched_and_fixed_seed}
\end{figure}}

The second experiment in Figure~\ref{fig:halton_and_random_batches} also shows the updated k-effective approximation after each transport sweep but compares iQMC's batch performance with the randomized-Halton (RQMC) samples and typical pseudo-random samples. While both sampling methods appear to hover about the true mean, the randomized-Halton samples have a much lower variance. Also noteworthy is that this test was run with over an order of magnitude fewer particles than the test in Figure~\ref{fig:batched_and_fixed_seed}. At this particle count, the batch method still produces a stable and accurate solution (unlike the fixed-seed method); however, it's notably less precise than when running with a higher particle count, indicating that more active batches are required to achieve the same level of precision in the converged solution. 
\textbf{\begin{figure}[ht]
  \centering
  \includegraphics[width=0.8\textwidth]{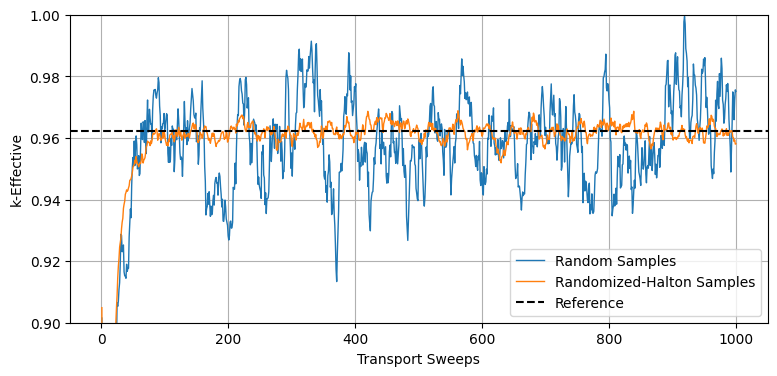}
  \caption{Updated k-Effective approximation after each transport sweep from the Takeda-1 problem with the iQMC batching with randomized-halton samples and pseudo-random samples. Each simulation was run using 1E5 particles per transport sweep.}
  \label{fig:halton_and_random_batches}
\end{figure}}

The third and final experiment shown in Figure~\ref{fig:error} shows the k-effective and mean scalar flux error from iQMC averaged across an ensemble of 20 simulations, each with a unique starting seed, as a function of the number of particle histories per transport sweep. Observing the convergence trends in Figures~\ref{fig:batched_and_fixed_seed} and~\ref{fig:halton_and_random_batches}, each simulation used 200 inactive and 100 active batches. Figure~\ref{fig:error} shows that although randomizing the QMC samples introduces some stochastic noise in the simulation, by averaging the converged solutions across $N_\text{active}$ batches this noise is reduced and very importantly the theoretical QMC convergence rate of $O(N^{-1})$ is maintained. The scalar flux convergence begins to plateau at higher particle counts and this is likely due to the coarse $25\times25\times25$ regular mesh used in each simulation. Increasing the fidelity of the mesh, particularly around areas of rapid flux change, would likely decrease this effect as was observed in previous iQMC studies~\cite{pasmann2024mitigating}.
\textbf{\begin{figure}[h!t]
  \centering
  \includegraphics[width=1.0\textwidth]{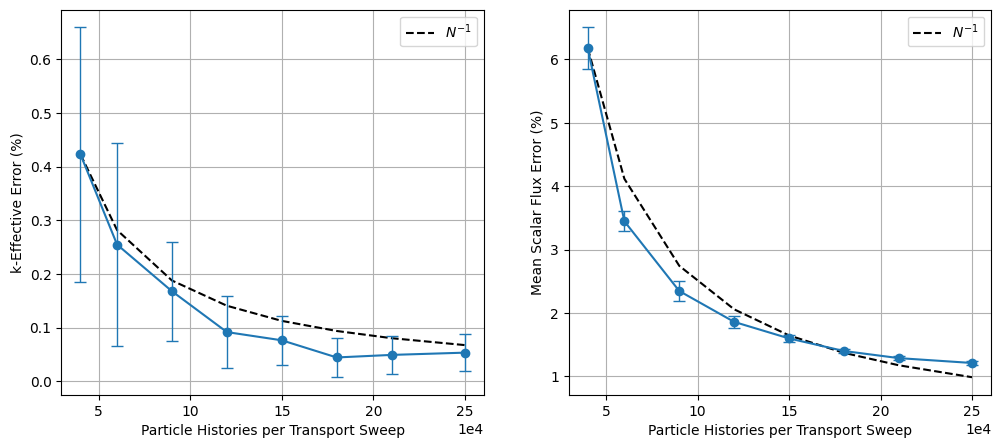}
  \caption{Batch iQMC solution convergence averaged across an ensemble of 20 simulations and shown with one standard deviation of the ensemble. Although the randomized-Halton samples introduce some statistical noise, the solutions still converge at the improved \textbf{$O\left(N^{-1}\right)$} rate.}
  \label{fig:error}
\end{figure}}

\section{CONCLUSIONS}\label{sec:conclusions}
In this analysis we have presented a new ``batch'' power iteration approach using randomized-Quasi-Monte Carlo samples in the iterative Quasi-Monte Carlo method for multigroup neutron transport simulations. The previously used fixed-seed approach introduced bias to the solution due to an under-sampling of the phase space similar to the ``ray effects'' observed in discrete ordinates methods. By randomizing the QMC samples each transport sweep with Owen's randomization, we gain a much better sampling coverage of the phase space and effectively remove this source of error. Crucially, despite introducing some stochastic noise to the solution, the improved $O(N^{-1})$ associated with QMC sampling was achieved. The rapid prototyping capabilities offered by the Monte Carlo Dynamic Code have led to significant algorithmic improvements to the iQMC method in recent years. This motivates future work of implementation in OpenMC~\cite{Romano2015} for more detailed performance assessments and comparisons to similar solvers like multigroup Monte Carlo and The Random Ray Method~\cite{Tramm2017}.

\section*{ACKNOWLEDGEMENTS}
This work was funded by Thea Energy and the Center for Exascale Monte-Carlo Neutron Transport (CEMeNT) a PSAAP-III project funded by the Department of Energy, grant number: DE-
NA003967 and the National Science Foundation, grant number DMS-1906446.

\bibliographystyle{mc2025}
\bibliography{main}

\end{document}